\documentclass[a4paper]{article}

\usepackage{amsmath,amssymb}
\newcommand{\STM}{\mathscr{Z}}
\newcommand{\SigmaS}{U}
\newcommand{\sigmaS}{u}
\newcommand{\Qs}{\mathbf{S}}
\newcommand{\qs}{s}

\newcommand{\NP}{\mathbf{NP}}
\newcommand{\PP}{\mathbf{PP}}
\newcommand{\BPP}{\mathbf{BPP}}
\newcommand{\ZPP}{\mathbf{ZPP}}
\newcommand{\Sat}{\mathtt{Sat}}
\newcommand{\MajSat}{\mathtt{MajSat}}
\newcommand{\sSat}{\sharp\mathtt{Sat}}

\newcommand{\PH}{\mathbf{PH}}
\newcommand{\CH}{\mathbf{CH}}

\newcommand{\Nat}{\mathbb{N}}

\usepackage[scr=boondoxo,scrscaled=1.05]{mathalfa}
\usepackage{chronology}

\newcommand{\longv}[1]{}

\title{On Randomized Computational Models and Complexity Classes: a Historical Overview}
\author{Melissa Antonelli \quad \quad Ugo Dal Lago \quad \quad Paolo Pistone}

\date{}

\begin{document}
\maketitle

\begin{abstract}
Since their appearance in the 1950s, computational models capable of performing probabilistic choices have received wide attention and are nowadays pervasive in almost every areas of computer science.
Their development was also inextricably linked with inquiries about computation power and resource issues.
Although most crucial notions in the field are well-known, the related terminology is sometimes imprecise or misleading.
The present work aims to clarify the core features and main differences between machines and classes developed in relation to randomized computation.
To do so, we compare the modern definitions with original ones, recalling the context in which they first appeared, and investigate the relations linking probabilistic and counting models.

\end{abstract}

\section{Introduction}
In the last few years, probabilistic computational models have been widely investigated, becoming pervasive in several fast growing areas of computer science (CS, for short).
Actually, the idea of relaxing the notion of algorithm from a \emph{deterministic} to a \emph{probabilistic} process appeared early in the history of modern computability, and randomized models started to be introduced already in the XX century~\cite{deLeeuw}.
Intuitively, a randomized algorithm is an algorithm involving some random process – typically corresponding to  ``flipping a coin'' – as part of its definition.
While in deterministic computation, for every input,
the algorithm produces (at most) one output, in the randomized case, the algorithm 
returns a set of outputs, each associated with the corresponding probability.

This peculiar feature makes probabilistic algorithms both efficient and powerful tools, with several applications in CS and technology. 
In some disciplines, like cryptography~\cite{GoldwasserMicali}, their use is not optional.
Today randomized programs and algorithms steer disciplines as robotics, verification and security, computer vision, cognitive science and NLP:

\begin{quote}
The last decade has witnessed a tremendous growth in the area of randomized algorithms.
During this period, randomized algorithms went from being a tool in computational
number theory to finding widespread applications in many types of algorithms.
Two benefits of randomization have spearheaded this growth:
simplicity and speed. \cite[p. ix]{MotwaniRaghavan}
\end{quote}
\normalsize
Generally speaking, randomized algorithms are crucial when dealing with uncertain information or partial knowledge, namely for all systems acting in realistic contexts – think, for example, of driverless cars~\cite{TBF} or of computer vision modelling~\cite{KollerFriedman}.

Historically, since the 1950s, various probabilistic abstract models have been introduced, from probabilistic Turing machines (PTMs, for short)~\cite{Santos69,Gill74} and $\lambda$-calculi~\cite{SahebDjaromi,JonesPlotkin}, to stochastic automata~\cite{Rabin63,Segala}.
Although these models are so relevant and wide-spread, it is not uncommon to find discrepancies in the corresponding terminology.
For instance, the terms \emph{counting} and \emph{threshold machines} are often used as interchangeable,  while it is not always considered that there exist multiple definitions of PTM; a PTM can  either be defined as a deterministic Turing machine (TM, for short) accessing a random-bit source in the form of an oracle-tape~\cite{deLeeuw,Gill74} or as a nondeterministic Turing machine (NTM) with transition functions to be chosen with (possibly equal~\cite{Gill77}) probability~\cite{Santos69}.
Sometimes these kinds of ambiguities make a precise comparison between works in the vast literature quite compelling and can even be the root of possible misunderstandings.
The aim of this paper is precisely that of clarifying fundamental notions in this area\footnote{This work focusses on \emph{proper} computational models only.
For reasons of space, we do not consider other relevant models, as MDP or Markov chains, and related probabilistic theories, e.g. probabilistic control.} and of investigating the differences between a few counting and probabilistic computational models,  the introduction of which was strongly linked to and influenced by the development of the corresponding complexity classes.

\paragraph{Structure of the Paper.}
Our work is bipartite. 
In Section~\ref{sec:machines}, we survey the historically-first definitions of probabilistic and counting machines.
In particular, in Section~\ref{sec:Early}, we present early probabilistic and stochastic machines by Davis~\cite{Davis}, Carlyle~\cite{Carlyle} and Rabin~\cite{Rabin63}, which are at the basis of the development of PTMs.
Then, in Section~\ref{sec:PTM}, original definitions of PTMs by Santos~\cite{Santos69} and Gill~\cite{Gill77} are presented and compared, showing that these models are not identical.
Finally, in Section~\ref{sec:threshold}, threshold~\cite{Simon} and counting machines~\cite{Valiant79} – which are often confused in subsequent literature – are inspected.
On the other hand, Section~\ref{sec:ProbCompl} is devoted to computational complexity.
In Section~\ref{sec:ProbClasses}, we focus on some relevant probabilistic and counting classes and show that they have been defined is inextricable relation with the corresponding machine models.
In Section~\ref{appCH}, we briefly consider the historically-first characterizations of $\CH$, as presented by Wagner~\cite{Wagner84,Wagner86} and T\'oran~\cite{Toran88}.
We conclude, in Section~\ref{sec:conclusion}, by summing up the main achievements of our work.

\section{Brief (Pre)History of Probabilistic and Counting Machines}\label{sec:machines}

\begin{sloppypar}
From the 1950s and 1960s on, probabilistic computational models started to receive attention~\cite{deLeeuw}, and machines including stochastic elements appeared.
In the 1970s, first formalizations of probabilistic and counting machines  were introduced \cite{Santos69,Gill74,Simon,Valiant79}.\footnote{Remarkably, in the same years, randomized $\lambda$-calculi were developed in the context of (probabilsitic) programming language theory, where a probabilistic program is basically one endowed with a (pseudo-)random number generator~\cite{JonesPlotkin,SahebDjaromi}.}
These new models were all explicitly defined as generalizations of 1960s standard (deterministic)  ones~\cite{Davis,Carlyle,Rabin63,Santos69,Santos69b}, but were also developed in mutual dialogue.
This even marked the beginning of a flourishing interplay between computational complexity and randomness.
\end{sloppypar}

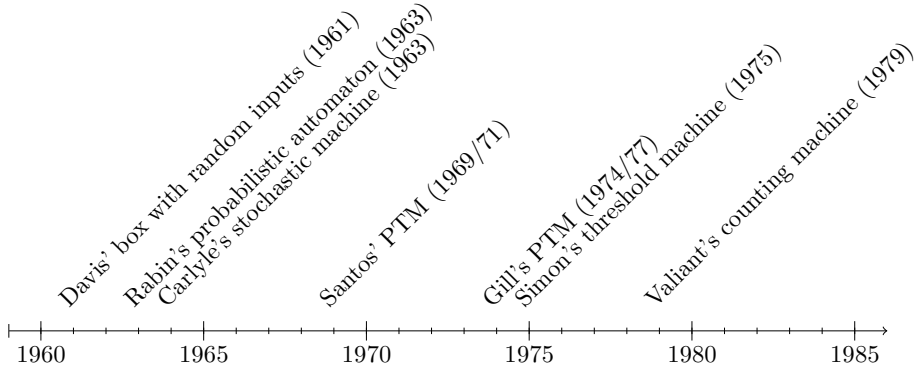
\begin{center}
\begin{chronology}{1959}{1985}{\textwidth}
\event[1961]{1961}{Davis' box with random inputs (1961)}
\event[1963]{1963}{Rabin's probabilistic automaton (1963)}
\event[1964]{1964}{Carlyle's stochastic machine (1963)}
\event[1969]{1969}{Santos' PTM (1969/71)}
\event[1974]{1974}{Gill's PTM (1974/77)}
\event[1975]{1975}{Simon's threshold machine (1975)}
\event[1979]{1979}{Valiant's counting machine (1979)}
\end{chronology}
\end{center}

\subsection{Prehistory: Early Probabilistic Machines}\label{sec:Early}
We start with a brief overview of three probabilistic machine models introduced in the 1960s, namely Davis' probabilistic automaton~\cite{Davis}, Carlyle' stochastic sequential machine~\cite{Carlyle} and Rabin's probabilistic automaton~\cite{Rabin63}.
All these models and works inspired Santos' and Gill's definitions of PTMs~\cite{Santos69,Gill77}.\footnote{For example,~\cite{Davis} is quoted by~\cite[p. 704, p. 706]{Santos69}, where also a comparison with~\cite{Carlyle} is provided,~\cite[p. 705]{Santos69}; \cite{Carlyle,Rabin63} are considered as the starting point for~\cite{Santos71}.
See also~\cite{SantosWee}.}

In 1961, Davis introduced his probabilistic automaton as an extension of the finite deterministic one~\cite[pp. 264-265]{Davis}.
His presentation started from the relation between Markov chains and automata theory.
After introducing the standard notions of finite automaton and finite Markov chain~\cite[p. 264]{Davis}, Davis defined probabilistic automata, called \emph{boxes with random input},\footnote{A box is a system $(X,\{\Gamma_\theta\})$, defining a finite automaton that consists of a set $X$ of states, together with a set $\{\Gamma_\theta\}$ of mappings $\Gamma_\theta:X\to X$ determining the behavior of the system for each value of the input parameter $\theta$.} and proved them to behave like Markov chains (and vice versa):

\begin{quote}
Consider a box whose consecutive input values are selected randomly and independently
by a sampling device of some kind.
The box $(X,\{\Gamma_\theta\})$ together with a probability distribution over the input values
will be called a \emph{box with random input}.
Its behavior is exactly that of a Markov chain.
More remarkably, however, is that the converse is also true.
\cite[p. 264]{Davis}
\end{quote}
\normalsize

In 1963, Carlyle formally presented his \emph{stochastic sequential machine}
as a 4-tuple including the \emph{conditional probability function}
$P(y;s' \; | \; x; s)$, which characterizes the joint probability that, given 
an input $x$ and a state $s$, the output is $y$ and the new state is 
$s'$~\cite{Carlyle}:\footnote{This machine is a generalization of Moore's deterministic one~\cite{Moore}, which, in Carlyle's terms, corresponds to the special case of a probability function taking values in $\{0,1\}$~\cite[p. 168]{Carlyle}.}

\begin{quote}
A \emph{stochastic sequential machine} $M=(X,Y,S,P)$
is defined through the specification of finite sets $X,Y,$ and $S$
(whose elements are called \emph{input symbols}, \emph{output symbols},
and \emph{states}, respectively),
together with a conditional probability function
$$
P(y;s' \; | \; s;x) : x \in X; y \in Y; s,s'\in S.
$$ 
The model $M$ is supposed to represent the idealized physical system with 
a finite number of distinct internal configurations
(states),
 such that (i) if input $x$ is applied when the state is $s$,
then $P(y;s' \; | \; s;x)$ is the joint probability that the observed response is
$y$ and the new state is $s'$ 
and (ii) if the machine is initially in state $s_1$ and inputs $x_1,x_2, \dots, x_n$ are applied successively, then the output sequence $y_1,y_2,\dots, y_n$ has the probability distribution:
$$
p_{s_1}(y_1y_2\dots y_n \; | \; x_1x_2 \dots x_n) = \sum_{s_k\in S, k\ge 2} \prod^n_{k=1} P(y_k ; s_{k+1} \; | \; s_k ; x_k).
$$
\cite[p. 167]{Carlyle}
\end{quote}
\normalsize
Then, the link with Markov chain is made explicit~\cite[p. 168]{Carlyle}.

In the same year, Rabin introduced his \emph{probabilistic automaton}.
In particular, in~\cite{Rabin63}, \emph{probabilistic automata} are defined 
by enriching  standard models with a \emph{probability transitions function},
which assigns a set of probabilities to each pair state-input:\footnote{Formally, Rabin's automata are 4-tuples $\langle S, M, s_0, F\rangle$, with $S$ set of states, $s_0\in S$ initial state, $F\subseteq S$ states over $\Sigma$, and $M:S\times \Sigma \to [0,1]^{n+1}$ probability transition  assigning a set of probabilities to each pair $(s,\sigma)$: when the machine in state $s$ receives $\sigma$ as input, it reaches any $s_i$ with a given probability, see~\cite[\textsc{Def.} 4, p. 234]{Rabin63}.}

\begin{quote}
It is quite natural to consider automata with stochastic behavior.
The idea is that the automaton when in state $s$ and when the input is $\sigma$, can move into any state $s_i$ and the probability for moving  into state $s_i$ is a function $p_i(s,\sigma)$ of $s$ and $\sigma$. [...] Though the generalization from the abstract deterministic automata to the abstract probabilistic automata (p.a.) lies near at hand, there are no general results about p.a. in the literature.
\cite[pp. 231-232]{Rabin63}
\end{quote}
\normalsize
As clear, these machines were again presented as natural generalizations of standard ones, this time of the automata defined in~\cite{RabinScott}.
In the subsequent pages, Rabin also developed a general theory of probabilistic automata.

\subsection{Probabilistic Turing Machines}\label{sec:PTM}
Roughly, a model for 
universal
computers is that of a machine enriched with random number generators.
Corresponding precise definitions were formalized in the form of PTMs in a series of 1970s papers by Santos~\cite{Santos69,Santos71} and Gill~\cite{Gill74,Gill77}, written in mutual dialogue with each other. 
A PTM is \emph{generally} defined as a machine that, at each computation step, chooses between two transition functions (usually) with equal probability $\frac{1}{2}$ and independently from previous choices, see e.g.~\cite{AroraBarak,Sipser}.
Yet, as we shall see, when looking at the original definitions by the mentioned authors, subtle but relevant differences emerge.

\subsubsection{Pioneering Probabilistic Machines}
The pioneering article \emph{Computability by Probabilistic Turing Machines}~\cite{deLeeuw} (1956) aimed at investigating the relation between probabilistic and deterministic machines in terms that can be summarized by the question: ``Is there anything that can be done by a machine with a random element but not by a deterministic machine?''~\cite[p. 185]{deLeeuw}.
The core of the paper was the attempt to 
make this inquiry less vague by introducing a class of probabilistic machines and by defining in a precise way the tasks they can perform.
It was then shown that, in such a specific and limited setting, the quoted question can be better (re-)formulated and a negative answer was given.
As we shall see,  this work was a fundamental stimulus and source of inspiration for many subsequent
studies on probabilistic machine models and complexity.

In particular, in this article probabilistic machines were intuitively defined as generalizations of standard machines, to be endowed with a random bit source.
More concretely, in the first Section~\cite[pp. 184ff.]{deLeeuw} computing machines were presented as machines with an infinite sequence of binary digits as inputs; in standard machines it is a fixed sequence (infinite tape), while for probabilistic machines it is the output of a binary random device, having probability $p$ of producing 1.
Otherwise said, a $p$-machine is a TM such that the (read-only) input sequence is randomly and independently chosen, so that each given digit in the sequence has probability $p$ of being 1 and probability 1 – $p$ of being 0:

\begin{quote}
We now wish to attach a random device to a machine in order to construct a probabilistic machine. 
If one has a device that prints 0's and 1's on a tape, 1's occurring with probability $p$ and 0's occurring with probability 1 – $p$ ($0<p<1$) and each printing being independent of the preceding printings, the output of the device can be used as the input tape of a machine.
The combination of the random device and the  machine will be called a $\underline{p\text{-machine}}$.
\cite[p. 188]{deLeeuw}
\end{quote}
\normalsize
As Fisher noticed, the (so-called) $p$-random input is equivalent to an oracle tape returning ``yes'' with probability $p$ on any given query~\cite[p. 481]{Fisher}.

On the other hand, ``doing more'' was defined in terms of set enumeration: the original question was re-stated in Theorem 2 (and, more formally, relying on the notion of $p$-enumerability, in~\cite[p. 191]{deLeeuw}), showing that if the $p$, defining the random device, is a  computable random number that can be enumerated by a $p$-machine (or even by a $\frac{1}{2}$-machine) of the type considered, then it can also be enumerated by a deterministic one.\footnote{Here, a real number is  computable when 
there is an
effective procedure for finding a digit of its binary expansion
and a set $S$ is $p$-enumerable if there is a $p$-machine such that $S=\big\{x$ $|$ Pr[$x$ is the output of $M$] $> \frac{1}{2}\big\}$.
If $p$ is a computable real number (or $\frac{1}{2}$) a machine with a random device cannot ``enumerate'' anything that a deterministic machine could not.
\longv{
such that $S=\{x$ $|$ Pr[$x$ is the output of $M$] $> \frac{1}{2}\}$.
If the binary expansion of $p$ is denoted by $A_p$,
then $S$ is $A_p$-enumerable if there is a machine $M$
such that $S$ is the output of the (deterministic) computation which results
when $A_p$ itself is placed on the input tape of $M$.
Then, the main theorem of the paper relates the three concepts of strong 
$p$-enumerability, $p$-enumerability,
and $A_p$-enumerability by showing them equivalent~\cite[p. 190]{deLeeuw}. 
So if $p$ is a computable real number (or $\frac{1}{2}$),
a machine with a random device cannot ``enumerate''
anything that a deterministic machine could not.}}
Finally, a more general class of machines, called computable-stochastic machines,
is defined~\cite[pp. 194ff.]{deLeeuw}.
\begin{quote}
A $\underline{stochastic \; machine}$ is to be an object having a countable set of states $X_1,X_2,X_3,\dots$, a distinguished initial state $X_0$, and a countable collection of output symbols $s_1,s_2,s_3,\dots$.
It is to be supplied with a rule that gives the probability that the machine will be next in state $X_p$, and print output symbol $s_q$ if it has passed successively through states $X_0,X_{j_1},\dots, X_{j_n}$ so far.
This probability will be denoted by $P(X_{j_1},\dots, X_{j_n}; X_p;s_q)$. \cite[p. 194]{deLeeuw}
\end{quote}
%
%
%
%
%
%
%
%
%
%
%
\subsubsection{On Santos' Probabilistic Turing Machine(s)}
Santos introduced his PTMs together with random functions in two papers: \emph{Probabilistic Turing Machines and Computability}~\cite{Santos69} (1969) and \emph{Computability by Probabilistic Turing Machines}~\cite{Santos71} (1971), where the 1969 machine is generalized.\footnote{Actually, for reasons of space, we will focus on the definition presented in~\cite{Santos68} only.
In~\cite{Santos71}, the 1969 PTM is called \emph{simple PTM} and represents a special case of a more general definition, from which it can be obtained by restricting one of the two defining conditions of the probability function. For further details on the generalized version of PTM defined by Santos in 1971, see~\cite[Sec. III]{Santos71}.}
These works were part of a more general study on probabilistic models, the author has developed in a series of 1960s-70s articles~\cite{Santos68,SantosWee,Santos69b,Santos69,Santos71}.

Santos started by informally presenting his PTMs as ``natural''~\cite[p. 165]{Santos71} generalizations of canonical TMs.\footnote{In particular, in~\cite{Santos69}, a formal definition of PTMs is presented in Sec. II. In Sec. III, Santos introduces computable random functions, and in Sec. IV he shows that the class of deterministically computable functions is a proper subclass of probabilistically computable ones.}
Then, in~\cite[Sec. II]{Santos69}, PTMs were defined more formally as tuples of the form $\STM=(\SigmaS, \Qs, p)$, where, as standard, $\SigmaS$ and $\Qs$ are the (nonempty) sets of symbols and states, respectively, but $p:\Qs\times U \times V \times \Qs \to [0,1]$ is the  ``new'' \emph{probability function}, for $V=\{\mathtt{R}, \mathtt{L}, \mathtt{T}\}$ being the set of symbols representing (resp.) right move, left move and termination.
Santos' probability function satisfies the following conditions:
\begin{itemize}
\itemsep0em
\item[i.]  for every $\qs \in \Qs$ and $\sigmaS\in \SigmaS$,
$\sum_{v\in V}\sum_{\qs\in \Qs}p(\qs, \sigmaS, v, \qs')=1$,
\item[ii.] for every $\sigmaS\in \SigmaS$, if $\qs\neq \qs'$, $p(\qs, \SigmaS, \mathtt{T}, \qs')
= 0$.
\end{itemize}
Intuitively, this function represents the \emph{conditional probability} for the ``next acts'' of $\STM$, given that it is at state $\qs$ and scanning a square on which $\sigmaS$ appears~\cite[p. 704]{Santos69}. 
In particular, condition i. ensures the existence of some next act.

Remarkably, this probability function \emph{fully} encapsulates the non-deterministic aspects of the machine.\footnote{For further details on the dynamics of Santos' PTMs, see~\cite[pp. 704--706]{Santos69}, where, in particular, $q_{\STM}(\alpha,\beta)$ denotes the \emph{probability} that, assuming the PTM $\STM$ has started with instantaneous expression $\alpha$, its next instantaneous expression is $\beta$.}
Observe also that, in this case, each next move of the machine can be associated with \emph{any} probability (satisfying conditions i.-ii.), not only with those corresponding to fair coin tosses.
Santos noticed that his PTMs behave like stochastic sequential machines~\cite{Carlyle}
and as the machines he defined in~\cite{Santos68,Santos69b}.
He also put them in relation with Markov chains.
Moreover, he remarks that standard TMs can be seen as special PTMs in which $p$ returns 0 or 1 only~\cite[p. 704]{Santos69}.

\subsubsection{On Gill's Probabilistic Turing Machines.}
A few years later, PTMs were also introduced by Gill in his 1972 Ph.D. thesis and in two  works, both called  \emph{Computational Complexity of Probabilistic Turing Machines}~\cite{Gill74,Gill77}.
One of the main motivations guiding Gill's interest in new probabilistic models was their deep connection with studies on randomized algorithms and functions, as first investigated in the seminal work by de Leeuw et al.~\cite{deLeeuw}:
\begin{quote}
Our goal is to contrast the computation resources – time or storage – of probabilistic versus  strictly deterministic algorithms'. \cite[p. 91]{Gill74}\footnote{The same sentence can be found in~\cite[p. 183]{deLeeuw}.}
\end{quote}
As probabilistic algorithms seem to lead to faster solutions for several problems, due to PTMs, Gill aimed at providing a \emph{formal model} to study them.

Indeed, in those years researchers realized  that certain types of problems appeared to be more easily or quickly solvable by randomized algorithms and tools, e.g.~via Monte Carlo simulations:

\begin{quote}
Recently certain problems have been shown to be solvable by probabilistic algorithms
that are faster than the known deterministic algorithms for solving these problems.
\cite[p. 675]{Gill77}
\end{quote}
\normalsize
\noindent 
As a concrete example,\footnote{Actually, he also considered Rabin probabilistic algorithm for finding the nearest pair of a set of $n$ points in $k$-space that executes in average time $O(n)$, and which is faster than the $n$ log $n$ steps required by the best known deterministic algorithms.} Gill explicitly mentioned probabilistic algorithms for primality testing by Rabin~\cite{Rabin76} and by Solovay and Strassen~\cite{SolovayStrassen}.\footnote{Remarkably, about his algorithm for primality test, Rabin wrote:
``
The salient features of our method are that it is \emph{probabilistic}, i.e.,~uses
randomization within the computation,
and that it produces the answer with a certain controllable miniscule probability 
of error.''~\cite[p. 129]{Rabin80}
%
%
When this algorithm determines a number composite, then its result is always true, but when it asserts that a number is prime, then there is a provably smaller probability of error.
More recently, a \emph{deterministic} poly-time algorithm for primality testing was presented in~\cite{AgrawalKayalSaxena}.
We thank Henri Stephanou for suggesting us to further analyze this example.}
In Section 4, he also introduced a new problem,\footnote{Actually, Gill himself noticed that, in this case, the speedup is limited and that the problem has no practical interest.} by presenting  a palindrome-like language that can be recognized by a  one-tape PTM to be \emph{faster} than any (one-tape) deterministic TM~\cite[\textsc{Th. 4.1}, p. 684]{Gill77}.
So, clearly, the development of his machines was guided by inquiries about their power and by connected resource issues:
Gill's definition of PTMs was intrinsically related to the study of new, probabilistic complexity classes (see Section~\ref{sec:ProbClasses}).

Intuitively, Gill's PTM is a TM  ``with the ability to flip coins in order to make random decisions''~\cite[p. 91]{Gill74}, so that its output is no longer uniquely determined.
A bit more formally, Gill's machines can be seen as (one-way, infinite tape) TMs associated with a \emph{random-source oracle} – in the form of a subroutine returning either 0 or 1 \emph{with equal probability} – and including a special \emph{coin-tossing state}.
When entering it, PTMs flip an unbiased coin and branch into one of two specified states, according to the given tossing output (and, as said, with equal probability):

\begin{quote}
\textsc{Definition 2.1.} A \emph{probabilistic Turing machine} (PTM) is a Turing machine with distinguished states called coin-tossing states.
For each coin-tossing state, the finite control unit specifies two possible next states.
The computation of the probabilistic Turing machine is determined except that in coin-tossing states the machine tosses an unbiased coin to decide between two possible next states. \cite[p. 676]{Gill77}\footnote{See also~\cite[p. 92]{Gill74}.}
\end{quote}
As for Santos' machines, the computation associated with Gill's PTMs is a stochastic process, this time driven by a random-bit supply corresponding to ``the simplest type of randomness''~\cite[p. 676]{Gill77}.

\subsubsection{Comparing Santos' and Gill's Machines}
In fact, there exist relevant differences between Santos' and Gill's PTMs.
First, the latter ones are presented as (deterministic) TMs accessing an oracle source of randomness, while the former are defined as NTMs with special transition functions returning a probability value.
Proving that oracle machines and machines accessing probability functions are analogous (even assuming that the source of randomness they access is equivalent, e.g.~corresponding to fair coin tossing) is not trivial, see e.g.~\cite{ADLP}.\footnote{Indeed, although it is more or less folklore to consider the oracle- and probability function-models as analogous, \emph{concretely} relevant dissimilarities depend on the way in which the machine accesses its source of randomness; for instance, showing the equivalence between the class of probabilistic recursive  functions $\mathcal{PR}$~\cite{DalLagoZuppiroli} – inspired by ``Santos-style PTMs'' and based on probabilistic functions – and of oracle recursive functions $\mathcal{OR}$ – based on the ``oracle'' model – is not straightforward, see~\cite{ADLP}.}
Second, in Gill's PTMs only unbiased choices are allowed, whereas the  ``kind of randomness'' defined by Santos is more general.
Indeed, the source of randomness considered in~\cite[\textsc{Def. 2.1}]{Gill77} can be seen as constituted by a sequence of i.i.d. bits:
%
\begin{quote}
The random inputs were required to be drawn from a source of
\emph{independent}, \emph{equiprobable} binary digits... \cite[p. 91]{Gill74}
\end{quote}
\normalsize
Gill himself was aware of this difference as he claimed that, by relaxing his model so to allow arbitrary bias, one rather obtains machines equivalent to Santos' ones, i.e.~in which transitions can be associated with any probability:
%
\begin{quote}
The more general probabilistic Turing machines of Santos can be simulated by coin-tossing machines, provided that coins with arbitrary biases are allowed. \cite[p. 678]{Gill77}
\end{quote}
\normalsize
He also presented his PTMs as a special case of Santos' one~\cite[p. 676]{Gill77}. 
Another relevant dissimilarity between the two machines is that, when entering a non-deterministic state, Gill's PTM chooses between two possible next states only, while Santos' transition function may lead to several subsequent configurations.

On the other hand, despite being more limited, Gill's PTMs – in which, as seen, the source of randomness is confined to the probability of independent and fair coin tosses – are somehow more realistic, as effectively implementable.
Nonetheless, both these original, alternative presentations have paved the way to \emph{different} definitions, \emph{equally} labelled as PTMs in the literature.

\subsection{Threshold and Counting Machines}\label{sec:threshold}
In the same decades, other machines related to probabilistic computation were introduced.
Generally speaking, the defining feature of both threshold and counting machines is their ability of enumerating or measuring the number of accepting computation paths.

\subsubsection{On Simon's Threshold Machine}
In 1975, Simon presented threshold machines in Chapter 4 of his Ph.D. thesis, \emph{On Some Central Problems in Computational Complexity}.
In particular, after introducing the notions of poly-time bounded NTMs and of computation trees in a formal way~\cite[pp. 88-89]{Simon}, in Section 4.4, he defined threshold machines as special NTMs accepting an input when the number of accepting paths is greater than a given threshold~\cite[pp. 90-91]{Simon}.
Actually, these machines are defined together with corresponding recognizable languages and having in mind the objective of extending $\NP$:

\begin{quote}
We say that given a tree (i.e. one node follows from the next by $M$'s
rules) the trees that satisfy 
$$
\exists \text{ an accepting  leaf}
$$
represent languages in NP.
The next question to ask is, what do trees S that are characterized by 
$$
(*) S\text{'s leaves satisfy predicate } R()
$$
[...] represent? [...] We may extend this question to: there are at least
$k$, no more than $k$,
at least half, etc.
The nice fact is that such machines recognizes exactly the $k$-NP
languages defined earlier.\footnote{The class $k$-$\mathbf{NP}$ is the class of languages recognized by a poly-time threshold machine. Notice that in~\cite[pp. 84ff.]{Simon}, Simon introduced a list of corresponding languages, including $k$-$\Sat$.
For further details, see Section~\ref{sec:ProbClasses}.} \cite[p. 90]{Simon}
\end{quote}
\normalsize
This shows, once again, how early developments of probabilistic and counting machines are intrinsically related with questions in complexity theory and to the definition of new probabilistic classes. 
%

\subsubsection{On Valiant's Counting Machines}
Counting machines were developed to analyze probabilistic and counting complexity as well.
In \emph{The Complexity of Computing the Permanent} (1979), Valiant introduced  them to define  the class $\sharp\mathbf{P}$.
Actually, he started by considering a specific problem – that of computing the permanent of an $n\times n$ matrix – and by claiming that it is at least as complex as counting the number of accepting computations of a poly-time NTM~\cite[p. 189]{Valiant79}.
It was to express (and prove) this statement in a precise way that Valiant introduced counting machines and the corresponding (function) class $\sharp \mathbf{P}$.\footnote{He also generalized these notions establishing a framework to classify counting problems and defining a corresponding hierarchy~\cite[p. 190]{Valiant79}.}

More precisely, in \cite[\textsc{Def. 2.1}, p. 191]{Valiant79}, a counting machine is defined as a (poly-time) NTM such that, given an input, it returns \emph{the number} of the corresponding accepting paths (in binary notation):
%
%
\begin{quote}
A \emph{counting Turing machine} is a standard nondeterministic TM with an auxiliary output device that (magically) prints in binary notation on a special tape the number of accepting computations induced by the input. \cite[p. 191]{Valiant79}
\end{quote}
\normalsize
Then, $\sharp\mathbf{P}$ is presented as the \emph{counting} class of functions that can be computed by a counting machine in polynomial time.
Finally, related notions of reduction and of complete problems are introduced.

\subsubsection{Comparing Simon's and Valiant's Machines}

Simon and Valiant themselves noticed the existence of mutual similarities
between their machines and with PTMs.
In~\cite[\textsc{Th.} 4.4, p. 91]{Simon}, Simon compared his machine model with Gill's  ones, showing that the class of languages recognized by poly-time threshold machines and by poly-time PTMs were the same.
On the other hand, Valiant noticed that threshold machines were very close – or ``essentially equivalent''~\cite[p. 190]{Valiant79} – to Gill's and Simon's ones.

Actually, as we have seen, although the two terms are often used as interchangeable in the literature, there is a crucial difference between counting and threshold machines: given an input, Simon's machine \emph{accepts} it when the number of accepting paths is greater than a given threshold, whereas Valiant's machine \emph{returns} the number (in binary) of the corresponding accepting paths.
Consequently, the class defined by Simon's machines, called $k$-$\mathbf{NP}$, is a class of problems, while $\sharp \mathbf{P}$ is a function class.

\section{On Probabilistic and Counting Classes}\label{sec:ProbCompl}
As said, the development of new probabilistic and counting models was inherently linked with  open questions in complexity theory and with the discovery of 
problems beyond the polynomial hierarchy ($\mathbf{PH}$, for short):

\begin{quote}
Since the notion of probabilistic machines
was introduced by Gill,
several researchers have been much interested
in several questions about its computational power.
\cite[p. 514]{Toda}
\end{quote}
\normalsize
This led not only to the definition of new probabilistic and counting classes, such as $\PP$ and $\sharp\mathbf{P}$, but also to the proof of their strong connection~\cite{Simon} and to the development of a counting hierarchy ($\mathbf{CH}$, for short), 
analogous to the polynomial one~\cite{ParberrySchnitger,Wagner86}.\footnote{Notably, $\PP$ constitutes the building block of the $\CH$, as defined by Wagner~\cite{Wagner86} and Tor\'an~\cite{Toran88}. For further details, see Section~\ref{appCH}.}

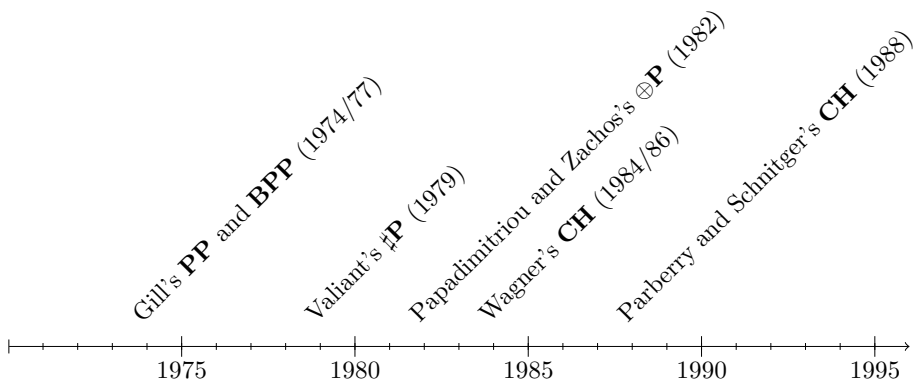
\begin{center}
\begin{chronology}{1970}{1995}{\textwidth}
\event[1974]{1974}{Gill's $\mathbf{PP}$ and $\mathbf{BPP}$ (1974/77)}
\event[1979]{1979}{Valiant's $\sharp \mathbf{P}$ (1979)}
\event[1982]{1982}{Papadimitriou and Zachos's $\oplus \mathbf{P}$ (1982)}
\event[1984]{1984}{Wagner's $\mathbf{CH}$ (1984/86)}
\event[1988]{1988}{Parberry and Schnitger's $\mathbf{CH}$ (1988)}
\end{chronology}
\end{center}

%
%
\subsection{On Probabilistic and Counting Classes}\label{sec:ProbClasses}

Crucial classes in the context of probabilistic and counting models are for example $\PP$ and $\sharp \mathbf{P}$.
In particular, $\PP$ can either be defined as the class of languages recognized by a poly-time PTM with an error probability smaller than $\frac{1}{2}$~\cite{Gill74,Gill77} or as the threshold class of languages recognized by a poly-time Simon's machine, accepting an input when the majority of its computation paths are accepting ones~\cite{Simon,Simon77}. 
In fact, in the latter case, this class was originally called $k$-$\NP$, but the two classes were shown equivalent in~\cite[\textsc{Th.} 4.4]{Simon}.
Specifically, in 1977, together with PTMs, Gill introduced some of the most popular classes to capture probabilistic (efficient) computation, namely $\PP$, $\BPP$ and $\ZPP$:

\begin{quote}
 \textsc{Definition 5.1} (i) $\PP$ is the class of languages recognized by polynomial bounded  PTMs. 
 (ii) $\BPP$ is the class of languages recognized by polynomial bounded PTMs with bounded error probability.
(iii) $\ZPP$ is the class of languages recognized by PTMs with polynomial bounded average run time and zero error probability.\footnote{Recall that, according to Gill, a PTM recognizes a language when it computes the characteristic function of the language and that a (partial) function $f$ computed by a PTM $M$ is defined by a function $f(x)$ returning $y$ when Pr$\{M(x) = y\}>\frac{1}{2}$; otherwise it is undefined.} \cite[p. 685]{Gill77}
\end{quote}
\normalsize
On the other hand, Simon defined the class \textbf{$k$-NP}~\cite{Simon} (or \textbf{$m$NP}~\cite{Simon77}) as that of languages recognized by a poly-time bounded threshold machine.
As anticipated, he also proved this class to be equivalent to that of languages recognized by a poly-time PTM in~\cite[\textsc{Th. 4.4}]{Simon} and in~\cite[\textsc{Th.} 2]{Simon77}, where he explicitly referred to Gill's restricted (viz., unbiased) machine as the probabilistic counterpart in his theorem(s). 
So, it emerges that distinct and independently-developed computational models have led to the introduction of the  \emph{``same'' class}, but also to \emph{different terminology}.

Both Gill and Simons defined complete problems for $\PP$ (resp., $k$-$\NP$).
In particular, in~\cite{Gill77}, Gill took into account $\MajSat$, which he actually called $\mathtt{Maj}$, and $\sSat$~\cite[\textsc{Df.} 5.7]{Gill77}.
$\MajSat$ was defined as the set of propositional formulas such that the majority of their interpretations are satisfying ones, whereas $\sSat$ is the set of pairs $\langle i,F\rangle$ such that the propositional formula $F$ has more than $i$ satisfying assignments.
On the other hand, Simon called this class $k$-$\Sat$~\cite[p. 83]{Simon} (or $m$-$\Sat$ in~\cite[p. 482]{Simon77}) and considered it together with other nineteen $k$-$\NP$ problems.
Remarkably, Gill explicitly mentioned Simon's completeness result~\cite[\textsc{Lemma 5.8}]{Gill77}  and extended it to $\MajSat$~\cite[\textsc{Prop. 5.10}]{Gill77}.

 As anticipated, a few years later the counting class $\sharp\mathbf{P}$ was introduced in~\cite{Valiant79}. 
 Valiant defined it as the class of functions computed by a counting machine in polynomial time~\cite[\textsc{Df.} 2.2]{Valiant79}, i.e.~as the class of functions returning \emph{the number of accepting paths} of a poly-time NTM.
Observe again that while $\PP$ is defined as a class of language~\cite{Gill77,Simon}, $\sharp \mathbf{P}$ is a function class.\footnote{Recall that, given an input, a threshold machines \emph{accepts} it when the number of corresponding accepting paths reaches the corresponding threshold, whereas a counting machine returns the \emph{number} of accepting paths.}
In the same years, Papadimitriou and Zachos studied the relationship between counting classes and $\PH$ and 
 introduced $\oplus \mathbf{P}$, the class of decision problems solvable by a poly-time NTM accepting
 an input when the number of its accepting paths is odd~\cite{PapadimitriouZachos}.\footnote{In 1985, Papadimitriou also defined a related, non-standard machine model called \emph{stochastic} machine, see~\cite{Papadimitriou85}.}
 Both these classes are strongly linked to $\PP$.\footnote{$\PP$ and $\oplus \mathbf{P}$ can be seen as the classes obtained by computing, respectively, the high-order and low-order bit of a $\sharp \mathbf{P}$ function.
 For further details, 
 see e.g.~\cite{BeigelGill}.}

\subsection{The Counting Hierarchy}\label{appCH}
 Since then, studies on probabilistic and counting complexity have spread 
and several definitions for such classes have been introduced.
It was in this context that, in the 1980s, $\CH$ was introduced for the first time both (and independently) by Wagner~\cite{Wagner86} and by Parberry and Schnitger~\cite{ParberrySchnitger}.
As for $\PH$, there exist (at least) two main, \emph{equivalent} presentations of this hierarchy:
Wagner's original one~\cite{Wagner84,Wagner86}, in terms of alternating quantifiers, and Tor\'an's oracle characterization~\cite{Toran88,Toran90}.
The latter is very similar to the corresponding polynomial version, but its building blocks are obtained by replacing $\NP$ with $\PP$.

In 1984/86, starting from the investigation of languages for succinct representation of combinatorial problems, Wagner introduced $\CH$ to classify natural problems in which counting is involved~\cite{Wagner84,Wagner86}.
His characterization relied on the notion of counting operator over classes of languages, $\textsf{C}$, which was inspired by the definition of threshold machine:
$\textsf{C}^{p(n)}_{f(x)}y P(y)$ expresses that there are at least $f(x)$ strings $y$ of length $p(n)$ satisfying $P$.
In particular, a clear presentation was offered in \emph{The Complexity of Combinatorial Problems with Succinct Input Representation} (1986), where, in Section 3, so-called $\mathscr{CPH}$ was introduced as an extension of $\PH$ obtained by adding $\textsf{C}$ to standard existential and universal quantifiers~\cite[p. 335]{Wagner86}.
Then, $\CH$ was defined as the smallest family of classes of languages including $\mathbf{P}$ and closed under existential, universal and counting operators.
For each level in the hierarchy, Wagner showed how to construct a complete set for it~\cite[p. 338]{Wagner86}.

Some years later, an alternative, but equivalent oracle characterization for $\CH$ was introduced by Tor\'an~\cite{Toran88,Toran90}.
First, he presented a slightly modified notion of $\textsf{C}$, the so-called \emph{exact counting quantifier}:

\begin{quote}
$\dots$for the function $f:\Sigma^* \to \Nat$, $f\in \mathbf{FP}$, a polynomial $p$, and a two argument predicate $P$, 
$$
\textsf{C}^p_{f(x)} y : P(x,y) \quad \Leftrightarrow \quad ||\{y : |y| \leq p(|x|) \text{ and }
P(x,y)\}||\ge f(x).
$$
We alternate now the polynomial counting quantifier $\textsf{C}$ with the existential and universal quantiiers in order to define the counting hierarchy.
\cite[p. 215]{Toran88}
\end{quote}
\normalsize
Then, in~\cite[Sec. 4]{Toran88}, the oracle characterization was introduced.
This definition is close to the oracle characterization of $\PH$, but here PTMs are used instead of NTMs~\cite[p. 219]{Toran88}:

\begin{quote}
One way to define the counting hierarchy is to take the usual definition of the polynomial hierarchy:
\begin{itemize}
\itemsep0em
\item $\Sigma^p_0=\mathbf{P}$ 
\item $\Sigma^p_{k+1}=\NP^{\Sigma^p_k}$ for $k\ge 0$
\end{itemize}
and replace ``$\NP$'' with ``$\PP$''. [...]
That gives us the definition:
\begin{itemize}
\itemsep0em
\item $\mathbf{C}^p_0=\mathbf{P}$
\item $\mathbf{C}^p_{k+1} = \mathbf{PP}^{\mathbf{C}^p_k}$ for $k\ge 0$.
\end{itemize}
Thus $\mathbf{C}^p_1=\PP$, $\mathbf{C}^p_2=\PP^{\PP}$, and so on.
\cite[pp. 182-183]{AllenderWagner}
\end{quote}
\noindent
Tor\'an also proved this characterization to be equivalent to Wagner's one.

Observe also that Wagner's counting operator was not the only ``probabilistic'' (class) quantifier that appeared in those decades to characterize complexity classes. 
For example, Papadimitriou introduced a \emph{probabilistic quantifier} to capture $\mathbf{PSPACE}$~\cite{Papadimitriou85}, Zachos and Heller defined \emph{random quantifiers} to characterize $\BPP$~\cite{ZachosHeller}, and, in~\cite{Zachos}, (again) Zachos considered the \emph{overwhelming} and \emph{majority} quantifiers.

\section{Conclusion}\label{sec:conclusion}
In this paper, we have surveyed some of the most relevant machine models and classes related to probabilistic and counting computation, which in turn is increasingly important in computer science.
We have shown that for several of these non-standard models more than one definition exists, and that the corresponding terminology is often ambiguous.
Such ambiguity is rooted in the genesis of these notions and in their historical development.
By inspecting historical sources, we have tried to identify the original features of the mentioned models so to avoid future misunderstanding and to enhance the comprehension of randomized computation.

For instance, we have shown that the very-first presentations of PTMs
are different – as said, PTMs are defined either as deterministic TMs accessing an oracle source of (unbiased) randomness or as NTMs with probabilistic transition functions
– and that the proof of their equivalence is not obvious (see~\cite{ADLP}). 
We have also seen that, although these terms are used more or less interchangeably in the literature, threshold  and counting machines  are not the same.
These differences also reflect on the way in which the corresponding complexity classes were defined. 
For example, we have clarified the distinction between the counting class $\sharp \mathbf{P}$, on the one hand, and the (equivalent) classes $\PP$, defined by Gill in relation with his PTM, and $k$-$\NP$, introduced by Simon relying on the threshold model, on the other.

\section{Acknowledgements}
The first author thanks Helsinki Institute for Information Technology (HIIT) for supporting her work since 2023.

%
%

\begin{thebibliography}{8}

\bibitem{AgrawalKayalSaxena}
Agrawal, M., Kayal, N. and Saxena, N.: Primes is in P, Ann. Math. \textbf{160}, 781--793 (2004)

\bibitem{ADLP}
Antonelli, M., Dal Lago, U. and Pistone, P.: On measure quantifiers in first-order arithmetic. In: Proc. CiE, pp. 12--24, 2021

\bibitem{AllenderWagner}
Allender, E. and Wagner, K.W.: Counting hierarchies: Polynomial time and constant depth circuit. Bull. EATCS \textbf{40}, 182-194.


\bibitem{AroraBarak}
Arora, S. and Barak, B.: Computational Complexity: A Modern Approach, Cambridge University Press, 2009


\bibitem{BeigelGill}
Beigel, R. and Gill, J.: Counting classes: threshold, parity, mods, and fewness. TCS, \textbf{103}, 3--23 (1992)


\bibitem{Carlyle}
Carlyle, J.M.: Reduced forms for stochastic sequential machines. J. Math. Anal. Appl. \textbf{7}, 167--174 (1963)


\bibitem{Davis}
Davis, A.S.: Markov chains as random input automata. Am. Math. Mon. \textbf{68}(3),  264--267 (1961)


\bibitem{DalLagoZuppiroli}
Dal Lago, U. and Zuppiroli, S.: Probabilistic recursion theory and implicit computational complexity. In: Proc. ICTAC, pp. 97--114, 2014


\bibitem{deLeeuw}
de Leeuw, K., Moore, E.F., Shannon, C.E. and Shapiro, N.: Computability by probabilistic Turing machines. Automata Studies,  \textbf{34}, 183--212 (1956)
 
 
 
\bibitem{Fisher}
Fisher, P.C.; Review: K. de Leeuw, E.F. Moore, C.E. Shannon, N. Shapiro, Computability by probabilistic machines, JSL, \textbf{35}(3), 481-482, 1970
 
 \bibitem{Gill74}
 Gill, J.T.: Computability complexity of probabilistic Turing machines. In: Proc. STOC,  pp. 91-95, 1974


\bibitem{Gill77}
Gill, J.T.:  Computability complexity of probabilistic Turing machines.
SIAM J. of Computing \textbf{6}(4), 675--695 (1977)


\bibitem{GoldwasserMicali}
Goldwasser, S. and Micali, S.: Probabilistic Encryption, J. Comput. Syst. Sci. \textbf{28}, pp. 279-299 (1984)

\bibitem{JonesPlotkin}
Jones, C. and Plotkin, G.: A probabilistic powerdomain for evaluations. In: Proc. LICS, 
pp. 186-195, 1989

\bibitem{KollerFriedman}
Koller, D. and Friedman, N.: Probabilistic Graphical Models: Principles and Techniques, MIT Press, 2009.


\bibitem{Papadimitriou85}
Papadimitriou, C.H.: Games against nature, JCSS, \textbf{31}(2), 288--301, 1985



\bibitem{PapadimitriouZachos}
Papadimitriou, C.H., and Zachos, S.K.: Two remarks on the power of counting. TCS \textbf{145}, 269--275, 1982


\bibitem{ParberrySchnitger}
Parberry, I., Schnitger, G.: Parallel computation with threshold functions. JCSS \textbf{36}, 278--302 (1988)


\bibitem{Rabin63}
Rabin, M.O.: Probabilistic automata. Inf. and Comput. \textbf{6}(3), 230--245 (1963)


\bibitem{Rabin76}
Rabin, M.O.: Probabilistic algorithms: In: Algorithms and Complexity,
pp. 21--39, 1976


\bibitem{Rabin80}
Rabin, M.O.: Probabilistic algorithms for testing primality. J. of Number Theory \textbf{12}, 128--138 (1980)


\bibitem{RabinScott}
Rabin, M.O., and Scott, D.: Finite automata and their decision problems.  IMB J. Res. Develop. \textbf{3}, 114--125 (1959)


\bibitem{SahebDjaromi}
Saheb-Djaromi, N.: Probabilistic LCF. In: Proc. MFCS, pp. 1--13, 1978


\bibitem{Santos68}
Santos, E.S.: Maximum automata. Inf. and Control, \textbf{13}, 363--377 (1968)

\bibitem{Santos69b}
Santos, E.S.: Maxmin sequential-like machines and chains. Math. Syst. Theory,
\textbf{3}(4), 300-309 (1969)


\bibitem{Santos69}
Santos, E.S.: Computability by probabilistic Turing machines. In: Proc. Am. Math. Soc., 
22(3)
pp. 704--710, 1969



\bibitem{Santos71}
Santos, E.S.: Computability by probabilistic Turing machines. Trans. Am. Math. Soc.,
\textbf{159}
159-165, 1971.



\bibitem{SantosWee}
Santos, E.S. and Wee, G.W.: General formulation of sequential machines. Inf. and Control, \textbf{12} 5-10, 1968.


\bibitem{Segala}
Segala, R.: A compositional trace-based semantics for probabilistic Automata.  In: Proc. CONCUR,  pp. 234--248, 1995


\bibitem{Simon}
Simon, J.: On some central problems in computational complexity. PhD thesis, Cornell University, 1975

\bibitem{Simon77}
Simon, J.: On the difference between one and many. In: Proc. ICALP, pp. 480--491, 1977

\bibitem{Sipser}
Sipser, M.: Introduction to the Theory of Computation, PWS, 2006 


\bibitem{SolovayStrassen}
Solovay, R., Strassen, V.: A fast Monte-Carlo test for primality,  SIAM J. Comput. \textbf{6}, 84--85 (1977)



\bibitem{Moore}
Moore, F.: Gedanken-experiments on sequential machines.  Automata Studies, \textbf{34}, 129--153 (1956)


\bibitem{MotwaniRaghavan}
Motwani, R., Raghavan, P.: Randomized Algorithms. Cambridge University Press, 1995


\bibitem{TBF}
Thrun, S., Burgard, W. and Fox, D.: Probabilistic Robotics, MIT Press, pp. 201-230.


\bibitem{Toda}
Toda, S.: On the computational power of PP and $\sharp$P. In: Proc. FOCS, pp. 514--519, 1989


\bibitem{Toran88}
Tor\'an, J., An oracle characterizatio of the counting hierarchy. In: Proc. Structures in Complexity Theory Third Annual Conference, pp. 213--223, 1988


\bibitem{Toran90}
Tor\'an, J., Complexity classes defined by counting quantifiers. J. of the ACM, \textbf{38}(3), 753--774 (1991)


\bibitem{Valiant79}
Valiant, L.G.: The complexity of computing the permanent. TCS, \textbf{8}(2), 189--201 (1979)


\bibitem{Wagner84}
Wagner, K.W.: Compact descriptions and the counting polynomial-time hierarchy. In: Frege Conference, 
pp. 383--392, 1984.

\bibitem{Wagner86}
Wagner, K.W.: The complexity of combinatorial problems with succinct input representations. Acta Inform., \textbf{23}, 325--356 (1986)


\bibitem{Zachos}
Zachos, S.W.: Probabilistic quantifiers and games, JCSS, \textbf{36}(3), 433--451 (1988)


\bibitem{ZachosHeller}
Zachos, S.W. and Heller, H.: A decisive characterization of BPP. Inf. and Control, 125--135 (1986).

\end{thebibliography}
%

\end{document}